# *Electromagnetic Momentum in a Dielectric: a Back to Basics Analysis of the Minkowski-Abraham Debate*

Joseph J. Bisognano*

*Department of Engineering Physics*
*University of Wisconsin-Madison*

*(Original Version: January 30, 2017; Revised Version: July 3, 2017)*

## Abstract

After more than a century of debate, there remains continuing discomfort over what is the correct expression for the electromagnetic momentum in a dielectric medium. This is the so-called the Minkowski-Abraham controversy. We show that there is indeed a consistent picture for the electromagnetic momentum associated with waves in a dielectric, but one must start with the fields *E* and *B,* not *D* and *H* as fundamental objects. The momentum flow is described by a tensor $T^{EB}$, where the expression is identical to that in vacuum. The Minkowski tensor and the Abraham force result from a fractured decomposition of the material momentum. However, for well localized fields, the Minkowski momentum is conserved, consistent with its identification as the canonical momentum. Several cases are discussed to demonstrate that this perspective is consistent and complete. In addition, it is shown that that care must be taken to include momentum transfer to bound charges at dielectric interfaces to correctly describe, for example, oblique reflections, where there is a normal component of the electric field. In an Addendum, alternate expressions for the local force on an electric dipole are discussed.

*Email: jjbisognano@wisc.edu*

Introduction

After more than a century of debate, there remains continuing discomfort over what is the correct expression for the electromagnetic momentum in a dielectric medium. Many articles concerning this so-called Minkowski-Abraham controversy have appeared over the decades, and many insights have been offered. See, for example [1,2].

However, an underlying unease persists. Even J. D. Jackson in his classic text [3] could only opine, "…although a treatment using the *macroscopic* Maxwell equations leads to an apparent electromagnetic momentum of $\boldsymbol{g} = \boldsymbol{D} \times \boldsymbol{B}$…the generally accepted expression for a medium at rest is $\boldsymbol{g} = \frac{1}{c^2} \boldsymbol{E} \times \boldsymbol{H}$...." In this note we make another attempt at clearing the air. In particular, we hope that this simple perspective will be of pedagogical value.

1. The Basic Dilemma

The macroscopic Maxwell equations lead to a simple expression [1,2] describing the change in mechanical momentum density of free charges in a dielectric medium:

$$\frac{d\boldsymbol{p}_{free}}{dt}\Big|_i = \{\rho_f \boldsymbol{E} + \boldsymbol{j}_f \times \boldsymbol{B}\}\Big|_i = -\frac{\partial \boldsymbol{g}^M}{\partial t}\Big|_i + \sum_{j=1}^3 \frac{\partial T_{ij}^M}{\partial x_j} \qquad (1.1)$$

where
$\boldsymbol{g}^M = \boldsymbol{D} \times \boldsymbol{B}$ is the Minkowski electromagnetic momentum density and $T_{ij}^M = E_i D_j + H_i B_j - \frac{1}{2}(\boldsymbol{E} \cdot \boldsymbol{D} + \boldsymbol{H} \cdot \boldsymbol{B})\delta_{ij}$ is the Minkowski stress tensor. The variables $\rho_f$ and $\boldsymbol{j}_f$ are the free charge and current densities, respectively, and *E, B, D, and H* carry their usual definitions. Integrating over an arbitrary volume, we have the statement that the rate of change of the mechanical momentum of free charges plus the Minkowski electromagnetic momentum equals the flow of Minkowski electromagnetic momentum through the volume's surface. In other words, by the divergence theorem, $(\overleftrightarrow{T^M} \cdot \boldsymbol{w})_i$ represents the flow of the $i^{th}$ component of momentum across a surface with normal $\boldsymbol{w}$. Similarly, the rate of change of the energy density is represented by the equation

$$-\boldsymbol{j}_f \cdot \boldsymbol{E} = \frac{\partial u}{\partial t} + \nabla \cdot \boldsymbol{S} \tag{1.2}$$

where $\boldsymbol{S} = \boldsymbol{E} \times \boldsymbol{H}$ represents the flow of energy through a boundary surface and $u = \frac{1}{2}[\boldsymbol{E} \cdot \boldsymbol{D} + \boldsymbol{B} \cdot \boldsymbol{H}]$ is the energy density.

Various problems arise in making equations (1.1) and (1.2) compatible. In particular, relativity requires for a particle, either massive or massless, that the velocity $v$, momentum $p$ and energy $E$ be related by the expression

$$p = Ev/c^2 \tag{1.3}$$

For a photon in a medium of dielectric constant $n$, $v=c/n$. Consider a transverse wave traveling in a homogeneous dielectric medium (with no magnetic polarization) of index of refraction $n$ propagating along the z-axis. Let the magnitude of the x component of the electric field be $E_0$. The y component of magnetic field $B$ is then $\frac{nE_0}{c}$ and $H = nc\varepsilon_0 E_0$. In terms of the tensor $\overleftrightarrow{T^M}$ and vector $S$, we have

$$|T^M{}_{zz}| = n^2 \varepsilon_0 E_0^2 \tag{1.4}$$

$$|S_z| = cn\, \varepsilon_0 E_0^2 \tag{1.5}$$

The expression relating momentum flow to energy flow (specifically, through a planar surface with normal in z-direction) analogous to Eqn 1.3 becomes

$$|T^M{}_{zz}| = |S_z|\frac{n}{c} = |S_z|\frac{1}{v} \tag{1.6}$$

Note that the velocity appears in the denominator, not the numerator, and appears in conflict with Eqn 1.3. Admittedly, in a dielectric the photons are "dressed" by the interaction in the material, and Eqn 1.3 may not be straightforwardly applied. If, however, the electromagnetic momentum density is given instead by the so-called Abraham expression $\boldsymbol{g}^A = \frac{1}{c^2}(\boldsymbol{E} \times \boldsymbol{H}) = \frac{1}{n^2}(\boldsymbol{D} \times \boldsymbol{B})$ (and, equivalently, the corresponding stress tensor is given by $\overleftrightarrow{T}^{\,reduced} = \frac{1}{n^2}\overleftrightarrow{T}^{\,M}$), we have

$$\left|T^{\,reduced}{}_{zz}\right| = |S_z|\frac{n}{n^2 c} = |S_z|\frac{1}{c^2}\frac{c}{n} = |S_z|\frac{1}{c^2}v \tag{1.7}$$

as desired by special relativistic considerations without any subtle argument. [Note: to avoid confusion, we denote the new tensor with the superscript "reduced" rather than A for Abraham, since there is another tensor, the symmetrized version of $T^M$, that appears in the literature associated with Abraham's argument.] This and other considerations such as angular momentum conservation and generalization to a fully covariant 4 dimensional symmetric stress tensor [1] lead to the debate on whether $\boldsymbol{g}^M$ or $\boldsymbol{g}^A$ best represents the electromagnetic momentum density.

2. Starting from Scratch

Our perspective is that $\boldsymbol{E}$ and $\boldsymbol{B}$ are the fundamental objects, and $\boldsymbol{D}$ and $\boldsymbol{H}$ are only convenient constructs. Moreover, arguments about Lorentz invariance using $\boldsymbol{D}$ and $\boldsymbol{H}$ are inherently suspect since the dielectric or magnetic materials imply a preferred frame of reference. So, we return to expressions for momentum density and flux in terms of $\boldsymbol{E}$ and $\boldsymbol{B}$. For the remainder of this paper we will consider only homogeneous dielectric materials, possibly with a planar discontinuity between two media (one may be vacuum). We assume there is no magnetization. Thus the discussion centers on $\boldsymbol{E}$ and $\boldsymbol{D}$, and $\mu_0\boldsymbol{H}=\boldsymbol{B}$. Our starting point is the expression for momentum conservation usually derived in a vacuum

$$\frac{d}{dt}(\boldsymbol{p_{mech}} + \varepsilon_0 \boldsymbol{E} \times \boldsymbol{B})|_i = \sum_j \frac{\partial}{\partial x_j} T_{ij}^{EB} \tag{2.1}$$

$$\frac{d}{dt}\left(\int_V (\boldsymbol{p_{mech}} + \varepsilon_0 \boldsymbol{E} \times \boldsymbol{B})\,d^3x\right)|_i = \sum_j \oint_S T_{ij}^{EB} w_j \,dS \tag{2.2}$$

where $\boldsymbol{w}$ is the outward normal to the closed surface S bounding volume V and

$$T_{ij}^{EB} = \varepsilon_0[E_iE_j + c^2B_iB_j - \tfrac{1}{2}(\boldsymbol{E}\cdot\boldsymbol{E} + c^2\boldsymbol{B}\cdot\boldsymbol{B})\delta_{ij}] \qquad (2.3)$$

<u>These expressions apply equally in a dielectric medium as long as the mechanical momentum density $\boldsymbol{p}_{mech}$ includes both the momentum carried by free charges and the bound charges in the dielectric medium.</u> Note that $\varepsilon_0 \boldsymbol{E}\times\boldsymbol{B} = \boldsymbol{g}^A$ when $\mu = \mu_0$. See Jackson [3], Chapter 6.7, for detailed derivation.

The point of all this is that Eqn 2.1-2.3 are the fundamental relations, and various expressions with $\boldsymbol{D}$ have hidden the momentum of the dielectric material. The tensor $T^{EB}$ represents the flow of electromagnetic momentum out of a volume and $\boldsymbol{p}_{mech}$ and $\varepsilon_0\boldsymbol{E}\times\boldsymbol{B}$ represent the total mechanical and electromagnetic momentum densities within the volume. We will argue that $T^{EB}$ is the object to be tracked in following momentum flow.

### 3. Sorting Out the Material Momentum

To make contact with the Minkowski expression, we must extract the material momentum change rate to derive an expression involving only the free charge momentum. We can mimic the derivation for Eqn 2.1-2.3, but only look at the momentum change of the bound charges and polarization. The bound charge density is given by

$$\rho_{bound} = -\nabla\cdot\boldsymbol{P} \qquad (3.1)$$

where $\boldsymbol{P}$ is the polarization and, for our simple system, $\boldsymbol{P} = \varepsilon_0(n^2-1)\boldsymbol{E}$. The constant $n$ is the index of refraction and $\boldsymbol{D}=n^2\varepsilon_0\boldsymbol{E}$. Similarly, since there is no magnetization,

$$\boldsymbol{j}_{bound} = \frac{\partial}{\partial t}\varepsilon_0(n^2-1)\boldsymbol{E} \qquad (3.2)$$

See Jackson [3], Chapter 6.6, for details. Then, we have for the material momentum $\boldsymbol{p}_{material}$

$$\frac{d\boldsymbol{p}_{material}}{dt} = \rho_{bound}\boldsymbol{E} + \boldsymbol{j}_{bound}\times\boldsymbol{B}$$

$$= -\varepsilon_0(n^2 - 1)[\mathbf{E}(\nabla \cdot \mathbf{E}) + (\mathbf{B} \times \frac{\partial}{\partial t}\mathbf{E})] \qquad (3.3)$$

Using $\nabla \times \mathbf{E} = -\partial \mathbf{B}/\partial t$, we have

$$\frac{d\mathbf{p}_{material}}{dt} = -\varepsilon_0(n^2 - 1)[\mathbf{E}(\nabla \cdot \mathbf{E}) - \mathbf{E} \times (\nabla \times \mathbf{E}) - \frac{\partial}{\partial t}(\mathbf{E} \times \mathbf{B})] \qquad (3.4)$$

Now, the first two terms in the bracket on the RHS are exactly the same terms that yield the electric field part of the stress tensor $T^{EB}$ of Eqn 2.3 in the usual vacuum derivation. We denote the resulting tensor as $T^E$. We have

$$\frac{d\mathbf{p}_{material}}{dt}\big|_i = -\varepsilon_0(n^2 - 1)[\sum_j \frac{1}{\varepsilon_0} \frac{\partial}{\partial x_j} T^E_{ij} - \frac{\partial}{\partial t}(\mathbf{E} \times \mathbf{B})\big|_i] \qquad (3.5)$$

Inserting into Eqn 2.1 yields

$$\left(\frac{d\mathbf{p}_f}{dt}\big|_i - \varepsilon_0(n^2 - 1)\left[\sum_j \frac{1}{\varepsilon_0}\frac{\partial}{\partial x_j} T^E_{ij} - \frac{\partial}{\partial t}(\mathbf{E} \times \mathbf{B})\big|_i\right] + \frac{\partial}{\partial t}(\varepsilon_0 \mathbf{E} \times \mathbf{B})\big|_i\right)$$

$$= \sum_j \frac{\partial}{\partial x_j} T^{EB}_{ij} \qquad (3.6)$$

To make contact with the Minkowski relation of Eqn 1.1, we note that the $-\varepsilon_0(n^2 - 1)\left[-\frac{\partial}{\partial t}(\mathbf{E} \times \mathbf{B})\right]$ can be combined with $\frac{\partial}{\partial t}(\varepsilon_0 \mathbf{E} \times \mathbf{B})$ to give $\frac{\partial}{\partial t}(\varepsilon_0 \mathbf{D} \times \mathbf{B})$. Likewise bringing the $T^E$ term to the RHS turns $T^{EB}$ into $T^M$ using $D = \varepsilon_0 n^2 E$. Thus Eqn 3.6 is identical to Eqn 1.1. We have that the Minkowski expression just has the material momentum change absorbed into the displacement current *D,* with some of the material momentum on the LHS and some on the RHS. There is nothing more to understanding the Minkowski expression than this fractured decomposition. However, this does not mean that the Minkowski momentum density has no physical meaning. When an electromagnetic wave packet is totally absorbed by a free particle in a finite region, the $T^E$ and the $T^{EB}$ vanish on the boundary. By the divergence theorem, their contribution to volume integrals of the momentum densities vanishes. This leaves the change of the absorbing free particle's momentum density determined by the term $\frac{\partial}{\partial t}(\varepsilon_0 \boldsymbol{D} \times \boldsymbol{B})$ alone as demanded by the Minkowski analysis. More generally, if one has well localized fields, with $T^E$ and the $T^{EB}$ vanishing sufficiently far away, it is the Minkowski momentum given by the integral of $\boldsymbol{g}^M$ that is conserved, consistent with its interpretation as the canonical momentum. However, if one is in a situation where there is momentum flow through boundaries such is in interfaces between different dielectric media, it is $T^{EB}$ alone that describes the total momentum flow. $T^E$ remains as part of the description of the momentum density carried by the material medium. One must remember that $\varepsilon_0 \boldsymbol{D} \times \boldsymbol{B}$ term includes both the field momentum and the material momentum, and it is misleading to consider it as the field momentum density.

To derive the Abraham version, the $T^E$ is again brought over the RHS to give $T^M$, with symmetrization not necessary for our simple homogeneous medium. On the LHS, the Abraham perspective does not combine the remaining material momentum and field momentum and leaves $\frac{\partial}{\partial t}(\varepsilon_0 \boldsymbol{E} \times \boldsymbol{B})$ alone as the "electromagnetic momentum density." The remaining piece of the material momentum density is noted as the "Abraham force density" $f^A$, a seemingly mysterious addition. Again, the decomposition hides the simplicity of Eqn 3.6.

4. Various Tensors

We have introduced a number of stress tensors, and here we would like to point them out. Consider a medium with no free charges.

a) Eqn 2.2 associates the time rate of change of the material momentum and electromagnetic momentum to the tensor $T^{EB}$, which is the fundamental object to describe momentum flow.

b) The Minkowski Eqn 1.1 implies that the time rate of change of volume integrated momentum density $g^M = \boldsymbol{D} \times \boldsymbol{B}$ is associated with the Minkowski tensor $T^M$; i.e.,

$$\frac{d}{dt}\left(\int_V \boldsymbol{D} \times \boldsymbol{B}\right) d^3x)|_i = \sum_j \oint_S T^M_{ij} w_j, \qquad (4.1)$$

but we have seen that this tensor combines electromagnetic momentum with a part of the material momentum.

c) Since $D = \varepsilon_0 n^2 E$, the Abraham momentum $\boldsymbol{g}^A = \varepsilon_0 \boldsymbol{E} \times \boldsymbol{B}$ *is* similarly associated with $T^{reduced} = T^M/n^2$ and

$$\frac{d}{dt}\left(\int_V \varepsilon_0 \boldsymbol{E} \times \boldsymbol{B}\right) d^3x)|_i = \frac{1}{n^2} \sum_j \oint_S T^M_{ij} w_j \qquad (4.2)$$

d) Using c), Eqn 3.5 can be rewritten as

$$\frac{d\boldsymbol{p}_{material}}{dt}\bigg|_i = \frac{n^2-1}{n^2} \sum_j \frac{\partial}{\partial x_j} T^B_{ij} \qquad (4.3)$$

where $T^B$ is the <u>magnetic</u> field part of $T^{EB}$ and also the magnetic field part of $T^M$. For a single travelling wave the electric and magnetic fields are proportional, but for situations with standing waves; e.g., at an interface or inside a non-reflective coating, the electric and magnetic fields are no longer proportional and using $T^B$ is crucial.

## 5. Momentum Conservation at Boundaries

Finally, we need to consider what happens at the boundary between two media of different index of refraction. Consider a simple planar boundary between two media (one possibly the vacuum, called A and B), say with normal in the z-direction. The flow of momentum through the boundary surface per unit area is given by $T^{EB}_{zz}$. Since we have assumed no magnetization, the parallel components of both the electric and magnetic field are continuous. This implies that for normal incidence of a transverse electromagnetic waves the flow of momentum into and out of the surface are equal. This situation, however, is different if the wave impinges obliquely at the interface. Then, there is a discontinuity in $T^{EB}_{zz}$ induced by the discontinuity in $E_z$ (a component normal to the interface) since it is $D_z$ that is continuous. From Eqn 2.3, the discontinuity in $T_{zz}^{EB}$ is given by

$$\Delta T\ _{zz}^{EB}=\ \tfrac{1}{2}\varepsilon_0[(E_z^B)^2-(E_z^A)^2\ ] \qquad (5.1)$$

$$\Delta T\ _{zz}^{EB}=\ [\varepsilon_0((E_z^B)-(E_z^A))\ ][\frac{(E_z^B)+(E_z^A)}{2}] \qquad (5.2)$$

This has a simple interpretation. The first bracket is just the material charge density driven by Gauss' law applied at the interface, "the tips of the dipoles making up the polarization at the discontinuity." The second bracket is just the average electric field seen by this charge density. We have that the discontinuity in $T^{EB}$, the change in momentum across the boundary, is just the momentum given to the material through this interaction. As we shall see, it is this extra interaction that is needed for a consistent picture of oblique reflections at interfaces such as two media or a perfectly conducting mirror in a medium. It removes, for example, the unphysical polarization dependence predicted in Peierls' argument analyzing reflection from a perfectly conducting mirror [2]. This, of course, is just a special case of a variable dielectric constant, which will be discussed briefly in our conclusions.

6. Examples

The starting point for these examples are standard solutions of Maxwell's equations for *E* and *B* obtained by matching boundary conditions. We only quote results since derivations can be found it many texts such as Jackson [3]. We then calculate $T^{EB}$ at interfaces to see if momentum and energy flow are consistent using $T^{EB}$ in various media.

6.1 Vacuum to Dielectric with Normal Incidence

Consider a wave coming along the *z*-axis normal to *x-y* plane. For *z<0* there is vacuum and for *z>0* there is a homogeneous dielectric medium of index of refraction *n*.

Let the incident wave be

$$\boldsymbol{E}_i = E_0 \hat{\boldsymbol{x}} \exp(ikz) \text{ and } \boldsymbol{B}_i = \frac{1}{c} E_0 \hat{\boldsymbol{y}} \exp(ikz) \tag{6.1}$$

with, in all cases, exp(-$i\omega t$) time dependence assumed.
Then, the reflected wave is given by

$$\boldsymbol{E}_r = \frac{1-n}{1+n} E_0 \hat{\boldsymbol{x}} \exp(-ikz) \text{ and } \boldsymbol{B}_r = -\frac{1}{c}\frac{1-n}{1+n} E_0 \hat{\boldsymbol{y}} \exp(ikz) \tag{6.2}$$

and the transmitted wave is given by

$$\boldsymbol{E}_t = \frac{2}{1+n} E_0 \hat{\boldsymbol{x}} \exp(ik'z) \text{ and } \boldsymbol{B}_t = \frac{n}{c}\frac{2}{1+n} E_0 \hat{\boldsymbol{y}} \exp(ik'z) \tag{6.3}$$

where $k = \frac{\omega}{c}$ and $k' = \frac{n\omega}{c}$.

These fields, of course, satisfy continuity conditions, namely

$$\boldsymbol{E}_i + \boldsymbol{E}_r = \boldsymbol{E}_t$$
$$\boldsymbol{B}_i + \boldsymbol{B}_r = \boldsymbol{B}_t$$

since there is no normal electric field. Since the interface surface normal is in the z-direction, the flow of z-momentum into the medium is given by $T_{zz}^{EB}$. From Eqn 2.3, we have at both sides of the boundary at z=0

$$T_{zz}^{EB} = -\tfrac{1}{4}\varepsilon_0 (\boldsymbol{E} \cdot \boldsymbol{E}^* + c^2 \boldsymbol{B} \cdot \boldsymbol{B}^*) = -E_0^2 \left(\frac{1+n^2}{(1+n)^2}\right) \tag{6.4}$$

For oblique reflection, there are two situations: the electric field normal and parallel to the plane of incidence. With the electric field normal, there is no $E_z$ and as discussed in Section 5, $T_{zz}^{EB}$ is continuous and equal on both sides of the boundary. When the electric field is parallel to the plane of incidence, there is now an $E_z$ and a discontinuity in $T_{zz}^{EB}$. However, this discontinuity, as given by Eqn 5.1 and 5.2, is just the momentum transferred to the material through the electric field interacting with the material surface charge induced by the polarization discontinuity.

*6.2 Vacuum to Dielectric to Vacuum with Normal Incidence and Thickness for No Reflection*

Now consider three regions separated by planes parallel to the x-y plane. Region A is vacuum from *z=-∞ to z=0*. Region B extends from *z=0* to *z=d*, and has an index of refraction *n*. Region C is vacuum and extends from *z=d* to *z=∞*. As in 6.1, consider a transverse wave coming from -∞ with x-polarization. If *d* is chosen for unity phase shift across region B there is no reflection and the electric fields are given at the AB and BC planes by

$$E_A \equiv E_0 \quad at\ AB\ in\ A$$

$$E_B^{forward} = E_0\ \frac{n+1}{2n} at\ AB\ and\ BC\ in\ B \tag{6.5}$$

$$E_B^{backward} = E_0\ \frac{n-1}{2n} at\ AB\ and\ BC\ in\ B \tag{6.6}$$

$$E_C = \alpha E_0 \quad at\ BC\ in\ C\ with\ |\alpha| = 1 \tag{6.7}$$

Then we have

$$T_{zz}^{EB} = \varepsilon_0 E_0^2\ on\ the\ A\ side\ of\ AB \tag{6.8}$$

$$T_{zz}^{EB} = \varepsilon_0 E_0^2\ on\ the\ C\ side\ of\ BC \tag{6.9}$$

$$T_{zz}^{EB} = \frac{1}{2}\varepsilon_0 E_0^2 \left[ \left(\frac{n+1}{2n} + \frac{n-1}{2n}\right)^2 + n^2 \left(\frac{n+1}{2n} - \frac{n-1}{2n}\right)^2 \right] = \varepsilon_0 E_0^2 \tag{6.10}$$

in region B at the AB and BC interfaces where the two middle terms are the electric and magnetic field contributions, respectively. Thus we have simple momentum conservation going through the layers. The Poynting vectors for the forward and backward momentum flux (±) are

$$\boldsymbol{S^\pm} = \boldsymbol{E} \times \boldsymbol{H} = \varepsilon_0 E_0^2 nc \left(\frac{n\pm 1}{2n}\right)^2 \tag{6.11}$$

Applying Eqn 4.2 and 4.3 to Eqn 6.10 yields
that the momentum flux in region B at both interfaces can be decomposed into a forward and backward EM momentum $\frac{1}{nc}S^{\pm}$ plus the expected material momentum from Eqn 4.3. Note that the flux associated both $S^+$ and $S^-$ are positive since positive momentum flow in the
positive direction and negative momentum flow in the negative direction have the same sign. Within the material there is a varying momentum flux that averages to the value at the interfaces and imply stress on the material.

*6.3 Vacuum to Dielectric of Index of Refraction n with Non-reflective Coating.*

Here we invoke a model of a non-reflective coating as a region of index of refraction $\sqrt{n}$ of proper length to cancel reflections and see the effect of standing waves in the coating. Consider three regions separated by planes parallel to the x-y plane. Region A is vacuum from *z=-∞ to z=0*. Region B extends from *z=0* to *z=d*, and has an index of refraction $\sqrt{n}$. Region C extends from *z=d* to *z=∞* and has an index of refraction *n.* As in 6.1, consider a transverse wave coming from -∞ with x-polarization. To act as a non-reflective coating *d* is chosen to give a phase shift *π/2* across region B. The electric fields are given at the AB and BC planes by

$E_A \equiv E_0 \quad at\ AB\ in\ A$

$$E_B^{forward} = E_0 \frac{\sqrt{n}+1}{2\sqrt{n}} at\ AB\ in\ B \tag{6.11}$$

$$E_B^{backward} = E_0 \frac{\sqrt{n}-1}{2\sqrt{n}} at\ AB\ in\ B \tag{6.12}$$

$$E_B^{forward} = iE_0 \frac{\sqrt{n}+1}{2\sqrt{n}} at\ BC\ in\ B \tag{6.13}$$

$$E_B^{backward} = -iE_0 \frac{\sqrt{n}-1}{2\sqrt{n}} at\ BC\ in\ B \tag{6.14}$$

$$E_C = i\alpha \frac{E_0}{\sqrt{n}} \text{ at BC in C with } \alpha = \exp(-in\frac{\omega}{c}z) \qquad (6.15)$$

Then we have

$$T_{zz}^{EB} = \varepsilon_0 E_0^2 \text{ on the A side of AB} \qquad (6.16)$$

$$T_{zz}^{EB} = \frac{1}{2}\varepsilon_0 E_0^2 \left(\frac{1}{n}+n\right) \text{ on the C side of BC} \qquad (6.17)$$

$$T_{zz}^{EB} = \frac{1}{2}\varepsilon_0 E_0^2 \left[\left(\frac{\sqrt{n}+1}{2\sqrt{n}}+\frac{\sqrt{n}-1}{2\sqrt{n}}\right)^2 + n\left(\frac{\sqrt{n}+1}{2\sqrt{n}}-\frac{\sqrt{n}-1}{2\sqrt{n}}\right)^2\right] = \varepsilon_0 E_0^2 \qquad (6.18)$$

on B side of AB interface.

$$T_{zz}^{EB} = \frac{1}{2}\varepsilon_0 E_0^2 \left[\left(\frac{\sqrt{n}+1}{2\sqrt{n}}-\frac{\sqrt{n}-1}{2\sqrt{n}}\right)^2 + n\left(\frac{\sqrt{n}+1}{2\sqrt{n}}+\frac{\sqrt{n}-1}{2\sqrt{n}}\right)^2\right]$$

$$= \frac{1}{2}\varepsilon_0 E_0^2 \left(\frac{1}{n}+n\right) \qquad (6.19)$$

on B side of BC interface.

Thus, we have at each interface momentum conservation. However, the momentum flow through AB is not equal to the momentum flow through BC. This implies that there is momentum deposition in the non-reflective coating. If one were to continue this to a second non-reflective coating (modeling an infinitely massive slab with non-reflective coatings on both sides inserted in a vacuum), the momentum deposition in the second coating would be opposite that of the first, with the outgoing wave carrying the same momentum in a vacuum as the incoming wave from the vacuum.

In region C, Eqn 6.17 can be re-expressed in terms of the energy flow $S_z = (\mathbf{E}\times\mathbf{H})_z = c\varepsilon_0 E_0^2$ in region C. Note that this implies no energy deposition in the non-reflective coating. We have

$$T_{zz}^{EB} = \frac{1}{2}\varepsilon_0 E_0^2 \left(\frac{1}{n}+n\right) = \frac{1}{2c}\left(\frac{1}{n}+n\right)S_z = \frac{1}{c}\left(\frac{1}{n}\right)S_z + \frac{1}{2c}\left(-\frac{1}{n}+n\right)S_z \qquad (6.20)$$

which matches Peierls' [2] expression for the total momentum flux as the average of the Minkowski and Abraham expressions and as the sum of Abraham field momentum flux and the material momentum flux. It is the momentum deposition in the non-reflective coating that yields overall momentum conservation. More generally, the Peierls' average applies when there is only a forward or backward wave, where $E$ and $B$ are proportional. In particular, this does not apply in the non-reflective coating. When there are both a forward and backward wave, the result moves toward the Minkowski expression, as we will see for a reflection on a mirror.

*6.4 Oblique Reflection on a Perfectly Conducting Mirror in Dielectric, both Normal and Parallel to Incident Plane Polarization*

We locate a perfectly conducting mirror on the *z=0* plane. Let the plane of incidence of a linearly polarized electromagnetic wave be the *x-z* plane. Consider an incident wave with the electric field normal to the plane of incidence and an angle of incidence θ. We have, for example, this set of consistent fields that includes both the incident and reflected waves

$$E_y = 2E_0 \sin(kz\cos\theta) \exp(-iky\sin\theta) \quad (6.20)$$

$$B_x = -2\frac{n}{c} E_0 \cos\theta \cos(kz\cos\theta) \exp(-iky\sin\theta) \quad (6.21)$$

$$B_z = 2E_0 \frac{n}{c} \sin\theta \sin(kz\cos\theta) \exp(-iky\sin\theta) \quad (6.22)$$

Note that at *z=0*, the only nonzero field is $B_x$.

So the averaged momentum flow at the *z=0* plane is

$$T_{zz}^{EB} = n^2 \varepsilon_0 E_0^2 \cos^2\theta \quad (6.23)$$

which is the accepted value of momentum transfer to a perfectly conducting mirror. Note that the momentum flow *n* scaling is consistent with the Minkowski expression since
the averaged incoming energy flux $S_z = n c \varepsilon_0 E_0^2 \cos\theta$. From sections 6.3 and 6.4, we see that the relationship between energy flow and momentum flow can differ in its "$n$" dependence. However, all cases are consistent with the $T^{EB}$ describing the momentum flow.

Now consider an incident wave with the electric field parallel to the plane of incidence and an angle of incidence $\theta$. We have, for example, this set of consistent fields that includes both the incident and reflective waves

$$E_x = 2E_0 \cos\theta \sin(kz\cos\theta) \exp(-iky\sin\theta) \tag{6.24}$$

$$E_z = 2E_0 \sin\theta \cos(kz\cos\theta) \exp(-iky\sin\theta) \tag{6.25}$$

$$B_y = 2E_0 \frac{n}{c} \cos(kz\cos\theta) \exp(-iky\sin\theta) \tag{6.26}$$

The averaged momentum flow at the $z=0$ in the dielectric is

$$T_{zz}^{EB} = -E_0^2 \varepsilon_0 (n^2 - \sin^2\theta) = -n^2 \varepsilon_0 E_0^2 \cos^2\theta - (n^2 - 1)\varepsilon_0 E_0^2 \sin^2\theta \tag{6.27}$$

which is unequal to Eqn 6.23 and seems to say that the mirror would receive a different momentum transfer dependent on whether the electric field is normal or parallel to the plane of incidence. However, referring back to Section 5 and observing that

$$-(n^2 - 1)E_0^2 \sin^2\theta = -(n^2 - 1)E_z^2 \tag{6.28}$$

we see that this extra term is just the momentum transferred to the dielectric medium through the bound charges from the "polarization dipole tips" generated by the electric field discontinuity. The remaining momentum that goes into the mirror proper is the same as Eqn 6.23. There is no polarization dependence on the momentum transfer to the mirror, but there is additional stress on the dielectric material at the interface.

7. Conclusion

We have shown that there is indeed a consistent picture for the electromagnetic momentum associated with waves in a dielectric. The starting point is to realize that $E$ and $B$, not $D$ and $H$ are the fundamental objects of analysis. The momentum flow is described by the tensor $T^{EB}$. In some situations, such as reflection by a mirror, the momentum flow $T^{EB}$ appears to be consistent with $\boldsymbol{g^M}$, while in others an average of $\boldsymbol{g^M}$ and $\boldsymbol{g^A}$ appears. But there really is no inconsistency if one keeps an eye on $T^{EB}$ as the fundamental object. In addition, care must be taken to include momentum transfer to bound charges at interfaces. Since we have not included magnetization, the differences between $H$ and $B$ have not been studied. It is expected that magnetization should be dealt with similarly, starting with Eqn 2.2 and 2.3. We have only considered situations with constant dielectric constant or a step discontinuity. The analysis can straight forwardly be extended to situations with a variable dielectric constant, again starting from Maxwell's equations and careful integrations by parts. The resulting expressions naturally develop $\nabla \varepsilon$ terms without having to accept the Minkowski, Abraham, or Einstein-Laub tensors, for example, as unproven starting points.

8. Acknowledgments

The author would like to thank Robert W. Boyd for the suggestion to look into this issue and for his comments on this note.

Appendix A. Forces in Dielectrics

In discussions of the forces in dielectrics two apparently conflicting expressions are commonly invoked: $-(\boldsymbol{\nabla} \cdot \boldsymbol{P})\boldsymbol{E}$ or $(\boldsymbol{P} \cdot \boldsymbol{\nabla})\boldsymbol{E}$. In Appendix A we offer our interpretation.

A1. The Issue

Barnett and Loudon [A1], as a starting point, describe two expressions for the electromagnetic force density in a dielectric, defining

$$\boldsymbol{f^c} = -(\boldsymbol{\nabla} \cdot \boldsymbol{P})\boldsymbol{E} + \dot{\boldsymbol{P}} \times \boldsymbol{B} \qquad (A1.1)$$

from an individual charge perspective and

$$f^d = (P \cdot \nabla)E + \dot{P} \times B \tag{A1.2}$$

from an individual dipole perspective, with polarization **P** in an electric field **E** and magnetic field **B**. Since the magnetic field terms are identical, we focus on the electric field terms. They consider a volume V which contains a dielectric and a thin volume of surrounding vacuum so the dielectric is fully contained in a region with zero polarization. Then they show by the divergence theorem that the integrated forces, **F**[c] and **F**[d] satisfy the relation

$$F_i^d - F_i^c = \int_V (P_j \nabla_j + (\nabla_j P_j))E_i dV = \int_V \nabla_j(P_j E_i) dV =$$

$$= \int_{surface} (P_j E_i) dS_j = 0 \tag{A1.3}$$

since the polarization is zero in the vacuum boundary layer.

A2. A Simple Example

Consider a simple dipole labeled "*n*" of equal plus and minus charges *q* (and for simplicity, mass) separated by a distance *2l*. Let the "center of mass" of the dipole be at position $x_n$, with the positive charge at position $x_n + l$ and the negative charge at $x_n - l$. Then the distribution of charge *ρ(x)* of a collection of such dipoles is given by

$$\rho(x) = q \sum_n [\delta(x - x_n - l) - \delta(x - x_n + l)] \tag{A2.1}$$

Note that for fixed **x**, the positive charge at **x** would derive from a dipole at center of mass position $x_n = x - l$ and the negative charge would derive from a dipole at center of mass position **x + l**. In other words, the positive and negative charges at fixed **x** are from different dipoles. See Kim [A2] for some clarifying diagrams.

To make contact with a smooth approximation of the polarization or dipole distribution, let *g(y)* be the distribution of the centers of mass of the collection of dipoles. Then the charge density is given by

$$\rho(x) = q \int d^3y \, g(y)\delta(x - y - l) - \delta(x - y + l)$$

$$= q[g(x - l) - g(x + l)] \qquad (A2.2)$$

which again shows that the charge density at *x* is a function of the dipole density at $x \pm l$. Identifying *g* with the polarization density *P* and using the relation

$$f(x) = \rho(x)E(x) \qquad (A2.3)$$

for the local force *f(x)* yields equation (A1.1).

A3. Forces on an Individual Dipole

Returning to the expression (A2.1), the force $f_n$ on a dipole whose center of mass is at $x_n$ is given by

$$f_n = q \int d^3x \, E(x)[\delta(x - x_n - l) - \delta(x - x_n + l)]$$

$$= q[E(x_n + l) - E(x_n - l)] \qquad (3.1)$$

This simply states that the force on an individual dipole is not localized at *x,* but at the location of the positive and negative charges. In particular, $f_n$ is not the local force at $x_n$. It is the force on a dipole whose center of mass is at $x_n$. Applying the derivative approximation to the difference yields expression (A1.2) which appears to yield a localized dipole force density *f(x)*. –But what it really says is that the force on a dipole whose center of mass is at *x* given by (A1.2). The force itself is distributed over the geometrical dimension of the individual dipoles. It does not represent the force density at *x.*

A4. Force on a Dielectric Slab

To summarize, equations (A1.1) and (A1.2) both correctly represent a portion of the force on a dielectric, but (A1.1) represents the force density at position *x*, whereas (A1.2) represents the distributed force density for dipoles whose centers of mass is localized at *x*. Clearly, when integrated over a volume containing the dielectric and a vacuum boundary layer, we get the same answer since all the charge, either place by place or dipole by dipole has been included. However, when the region does not have this vacuum layer, results can be different, with the discrepancy coming from boundary charges "sticking out" of the integration volume. For example, the dipoles whose centers of mass are just within a boundary actually have charge outside the boundary— the surface charge—but are part of the integral of equation (A1.2) over a volume containing the dipole centers of mass. One might want to argue that the distance is infinitesimal, but that is what the notions of surface charge and $\delta$-functions are addressing. Counting the force of those boundary charges as being within the boundary can be confusing. Either expression may have applicability, depending on what means by measuring the force in a volume of dielectric. For example, since a given dipole is tightly bound, the force on the positive and negative components of the dipole could act in together on an underlying structure within the volume even if some of the dipole charge was outside the integration volume.